\documentclass[11pt,amstex,amssymb,titlepage,a4paper]{article}
\usepackage{epsfig}
\usepackage{amsmath}
\usepackage[english]{babel}

\newcommand{\D}{\partial}
\newcommand{\mkmath}[1]{\ifmmode{#1}\else{\mbox{$#1$}}\fi}
\newcommand{\mb}[1]{\mkmath{\mathbf{#1}}}

\begin{document}
%---------------------------------------------------------------------------
\begin{titlepage}

\begin{center}

{\large\bf Static and Time Dependent Density Functional\\[.2cm] 
Theory with Internal
  Degrees of Freedom: Merits and \\[.2cm]Limitations Demonstrated for the
  Potts Model
} 

\vspace{1cm} S.~Heinrichs\footnote{Fachbereich Physik, Universit\"at
  Konstanz, 78457 Konstanz, Germany}, W.~Dieterich$^1$,
P.~Maass\footnote{Institut f\"ur Physik, Technische Universit\"at
  Ilmenau, 98684 Ilmenau, Germany} and
H.~L.~Frisch\footnote{Department of Chemistry, SUNY at Albany, Albany,
  New York 12222, U.S.A.}

\vspace{1cm}
7. April 2003

\vspace{1cm}
\begin{minipage}[t]{11cm}
  \begin{center}
  {\bf Abstract}
\end{center}

We present an extension of the density-functional theory (DFT)
  formalism for lattice gases to systems with internal degrees of
  freedom. In order to test approximations commonly used in DFT
  approaches, we investigate the statics and dynamics of occupation
  (density) profiles in the one-dimensional Potts model. In
  particular, by taking the exact functional for this model we can
  directly evaluate the quality of the local equilibrium approximation
  used in time-dependent density-functional theory (TDFT). Excellent
  agreement is found in comparison with Monte Carlo simulations.
  Finally, principal limitations of TDFT are demonstrated.
  
  \vspace{.4cm}
  Keywords: density functional theory, local equilibrium\\
  \phantom{Keywords: }approximation, non-equilibrium dynamics,\\
 \phantom{Keywords: }potts model, lattice gas kinetics
\end{minipage}

\end{center}

\end{titlepage}
\pagenumbering{arabic}          % Arabic numbers from here on

%------------------------------------------------------------------------------
\section{Introduction}
%------------------------------------------------------------------------------

Density functional theory (DFT) and its time-dependent variant (TDFT)
are powerful methods to derive phase diagrams and the kinetics of
phase transformations \cite{Loewen:1993,Evans:1979,Gouyet/etal:2003}
in condensed matter systems, in particular in the presence of
confinement effects. Particularly useful are these theories for
lattice systems \cite{Nieswand/etal:1993,Reinel/Dieterich:1996},
allowing us to deal with the discrete nature of structures encountered
e.g.\ in the description of metallic alloys, adsorbate layers, and
complex pattern formation on atomic scales. 

In the case where many ordered phases can coexist one is lead to
include internal degrees of freedom into the DFT approach. As a
prominent model we will consider in this work the $q$-state Potts
model, which has a $q$-fold degenerate ground state \cite{Potts:1952}.
The non-equilibrium dynamics of that model reflects the coarsening of
domains following a quench from the disordered homogeneous phase to a
system with long--range order seen in binary alloys, liquid crystals,
magnetic bubbles, Langmuir films and soap bubbles
\cite{Sire/Majumdar:1995}. The Potts model (in the limit $q \to 1$) is
isomorphic to a site-bond percolation problem
\cite{Kasteleyn/Fortuin:1969,Coniglio/Klein:1980} and for $q=2$ it
corresponds to the Ising model \cite{Chaikin/Lubensky:1995}. There are
interesting experimental realizations for $q=3$ (e.g.\ Kr on Graphite
\cite{Birgeneau/Horn:1986}), $q=4$ and $q=\infty$
\cite{Glazier/etal:1989} (froths and metallic grains).

In this paper, based on the exact density functional for the
one--dimensional $q$-state Potts model \cite{Buschle/etal:2000}, we
investigate the quality of various approximations often employed in
DFT and TDFT. After presenting a general scheme for treating systems
with internal degrees of freedom, we first consider the mean spherical
approximation (MSA) for the equilibrium properties, and show how it
compares with a mean-field approach and the exact solution. We then
derive the TDFT \cite{comm} for the one--dimensional Potts model and,
as an application, study the smoothening of an initial sharp-kink
density profile. Let us note that this TDFT differs in many respects
from the Runge-Gross theory \cite{Runge/Gross:1984} for electronic
systems in a time-varying external field, one major difference being
that the present approach assumes stochastic, overdamped dynamics. The
basic techniques to set up a TDFT for classical systems are reviewed
in \cite{Marconi/Tarazona:1999,Gouyet/etal:2003}. By comparison with
Monte--Carlo simulations we show that the TDFT provides a significant
improvement over kinetic mean field theory. Moreover, since in our
case the TDFT is based on an exact density functional, the differences
between the simulated and DFT results allow us to perform a specific
test of the local equilibrium approximation used in the TDFT\@.
Finally, we discuss principal limitations of the TDFT, where only the
density profile is used to specify a non-equilibrium state of the
system. This amounts to an incorrect account of correlations.

%------------------------------------------------------------------------------
\section{DFT for Lattice Gases with Internal Degrees of Freedom}
%------------------------------------------------------------------------------
We consider a lattice gas where each site $i$ is either vacant
($x_i^{\alpha}=0$ for all $\alpha$) or singly occupied by a particle
in one of $q$ internal states (note that this corresponds to a
generalized Potts model with $q+1$ states). If state $\alpha \in
\{1,\dots,q\}$ is realized, $x_i^{\alpha}=1$ and $x_i^{\beta}=0$ for
$\beta\ne \alpha$. Occupation numbers therefore satisfy
$x_i^{\alpha}x_i^{\beta}=x_i^{\alpha}\delta_{\alpha\beta}$. The
Hamiltonian including two particle interactions
$\Phi_{i,j}^{\alpha\beta}$ and site energies $\epsilon_i^{\alpha}$ due
to an external potential is given by
\begin{equation}
  \label{hamilton1}
  H = \frac{1}{2}\sum_{i\ne j}
\sum_{\alpha,\beta}\Phi_{i,j}^{\alpha\beta} x_i^{\alpha}x_j^{\beta}
+\sum_{i,\alpha}\epsilon_i^{\alpha}x_i^{\alpha}
\end{equation}
The formal steps of DFT for fluids or lattice gases without internal
degrees of freedom can be carried over to the case considered here.
One arrives at a variational principle based on the functional for the
grand canonical potential
\begin{equation}
  \Omega[\mb{p}]=
F[\mb{p}] + \sum_{i,\alpha}\tilde\epsilon_i^{\alpha}p_i^{\alpha}
\label{eq:omega}
\end{equation}
where $\tilde\epsilon_i^{\alpha}=\epsilon_i^{\alpha}-\mu_\alpha$,
$\mu_\alpha$ being the chemical potential fixing the mean total
occupation $\bar p_\alpha$ of state $\alpha$,
$\mb{p}=\{p_i^{\alpha}\}$ and $p_i^{\alpha}=\left< x_i^{\alpha} \right>$
are the average occupation numbers. The free energy functional is
decomposed into an 'ideal' part describing a non-interacting lattice
gas,
\begin{equation}
  F_{\rm id}[\mb{p}]=\Biggl[ \sum_{i,\alpha}p_i^{\alpha} \ln
  p_i^{\alpha} + \sum_i\left(1-\sum_{\alpha}p_i^{\alpha} \right) \ln
  \left(1-\sum_{\alpha}p_i^{\alpha} \right) \Biggr]
\end{equation}
and an excess part $F_{\rm ex}[\mb{p}]$ due to interactions (for
convenience we set $k_BT=1$). The equilibrium occupation is then
obtained by minimizing $\Omega[\mb{p}]$ with respect to the
$p_i^{\alpha}$. The corresponding equations $\D\Omega/\D
p_i^{\alpha}=0$ determining the equilibrium profile are called
structure equations. The minimum value of $\Omega[\mb{p}]$ is the
grand-canonical potential at equilibrium.
 
Higher order derivatives of $\Omega[\mb{p}]$ with respect to the
$p_i^{\alpha}$ taken at the equilibrium profile yield a hierarchy of
direct correlation functions  \cite{Evans:1979}.
In particular, by differentiating the structure equation, the
inhomogeneous Ornstein--Zernike equation
\begin{equation}
  \tilde{c}_{ij}^{\alpha\beta}+\sum_{k,\gamma}\tilde{c}_{ik}^{\alpha\gamma}
  \; p_{k}^{\gamma} \; h_{kj}^{\gamma\beta} = h_{ij}^{\alpha\beta}
\label{eq:oz}
\end{equation}
can be derived \cite{Nieswand/etal:1993}, which relates the direct
correlation function $c^{(2)}_{i\alpha,j\beta}=-\D^2 F_{\rm ex}[\mb{p}]/\D
p_i^{\alpha}\D p_j^{\beta}$ entering $\tilde{c}_{ij}^{\alpha\beta} =
c^{(2)}_{i\alpha,j\beta}-\delta_{ij}/[1-\sum_{\mu}p_i^{\mu}]$ to the
pair correlation function
$h_{ij}^{\alpha\beta}=g_{ij}^{\alpha\beta}-1=(1\!-\!\delta_{ij})
\langle x_i^{\alpha} x_j^{\beta} \rangle/\langle x_{i}^{\alpha}\rangle
\langle x_{j}^{\beta}\rangle$. In the limit $q=1$ the known result for
systems without internal degrees of freedom is recovered.

%-------------------------------------------------------------------------------
\section{Application to the 1D Potts model}
%-------------------------------------------------------------------------------
In the following, standard approximation schemes used in DFT are
tested based on the one-dimensional Hamiltonian ($1\le i\le M$)
\begin{equation}
  \label{hamilton2}
  H = \sum_{i,\alpha,\beta}v_i^{\alpha\beta}{x_i^{\alpha}x_{i+1}^{\beta}} 
+ \sum_{i,\alpha}{\tilde\epsilon_i^{\alpha}x_i^{\alpha}}\,,
\end{equation}
which for $v_i^{\alpha\beta}=V\delta_{\alpha\beta}$ reduces to
the standard Potts model \cite{Potts:1952}.

\subsection{Exact density functional}
For one-dimensional lattice gases with short range interactions, a
general scheme for deriving exact density functionals based on Markov
chains has recently been developed in \cite{Buschle/etal:2000}. This
approach in particular provides an exact functional for the Potts
model, which has been derived earlier by Percus \cite{Percus:1982}.
The functional can be written as
\begin{eqnarray}
\Omega[\mb{p}]&=&\sum_{i=1}^{M}\Biggl\{
\sum_{\alpha=1}^q \tilde\epsilon_i^{\alpha} p_i^{\alpha}+
\sum_{\alpha,\beta=1}^q v_i^{\alpha\beta}\Gamma_i^{\alpha\beta}+
\nonumber\\
&&\phantom{\sum_{i=1}^M\Biggl\{}\hspace*{-0.5cm}+
\sum_{\beta} \left[
   \sum_{\alpha} \Gamma_i^{\alpha\beta}
  \log\frac{\Gamma_i^{\alpha\beta}}{p_{i-1}^{\beta}}+ 
    \Bigl(p_{i-1}^{\beta}\!-\!\sum_{\alpha} \Gamma_i^{\alpha\beta}\Bigr)
      \log\Bigl(1\!-\!\frac{\sum_{\alpha} \Gamma_i^{\alpha\beta}}
         {p_{i-1}^{\beta}}\Bigr)\right]\nonumber\\
&&\phantom{\sum_{i=1}^M\Biggl\{}\hspace*{-0.5cm}+
\sum_{\alpha}\Bigl(p_i^{\alpha}\!-\!\sum_{\beta}
\Gamma_i^{\alpha\beta}\Bigr)\log\Bigl(\frac{p_i^{\alpha}\!-\!\sum_{\beta}
\Gamma_i^{\alpha\beta}}{1-\sum_{\beta} p_{i-1}^{\beta}}\Bigr)\\
&&\phantom{\sum_{i=1}^M\Biggl\{}\hspace*{-0.5cm}+
\Bigl(1\!-\sum_{\beta} (p_{i-1}^{\beta}\!+\!p_i^{\beta})\!+\!\sum_{\alpha,\beta} 
\Gamma_i^{\alpha\beta}\Bigr)\log\Bigl(1\!-\!
\frac{\sum_{\beta} p_i^{\beta}\!-\!\sum_{\alpha,\beta} \Gamma_i^{\alpha\beta}}
{1\!-\!\sum_{\beta} p_{i-1}^{\beta}}\Bigr)
\Biggr\}\,,\nonumber
\label{pottsfunc-eq}
\end{eqnarray}
where the correlators 
\begin{equation}
\Gamma_i^{\alpha\beta}=\langle
x_{i-1}^{\alpha} x_i^{\beta} \rangle
\end{equation}
have to be expressed by the mean occupation numbers $\{p_{k}^{\gamma}\}$.
This is achieved by solving the correlator equations
\cite{Buschle/etal:2000}
\begin{equation}
\Gamma_i^{\alpha\beta}=e^{-v_i^{\alpha\beta}}\frac{(p_i^{\alpha}-
\sum_{\gamma} \Gamma_i^{\alpha\gamma})(p_{i-1}^{\beta}
-\sum_{\delta} \Gamma_i^{\delta\beta})}
{1-\sum_\gamma (p_{i-1}^{\gamma}+p_i^{\gamma})+
\sum_{\gamma,\delta} \Gamma_i^{\delta\gamma}}\,.
\label{pottscorrel-eq}
\end{equation}
A simpler expression for the functional is obtained if the vacancies
are considered as an additional Potts state with index $\alpha=0$ (and
$v_i^{\beta0}=v_i^{0\beta}=0$), whereby the lengthy entropic part
reduces to $\sum_{i=1}^M \sum_{\alpha,\beta=0}^q
\Gamma_i^{\alpha\beta} \log(\Gamma_i^{\alpha\beta}/p_i^{\alpha})$. In
the fully occupied case ($\sum_\alpha x_i^{\alpha}=1$) the solution of
eqs.~(\ref{pottscorrel-eq}) requires a careful limiting procedure by
letting the vacancy concentration go to zero.

The structure equations read
\begin{equation}
e^{-\tilde \epsilon_i^{\alpha}}=\frac{\displaystyle (1-\sum_{\beta}p_i^{\beta})\left(p_i^{\alpha}-
\sum_{\gamma} \Gamma_i^{\alpha\gamma}
\right)\left(p_i^{\alpha}-\sum_{\delta} \Gamma_{i+1}^{\delta\alpha}\right)\frac{1}{p_i^{\alpha}}}
{\displaystyle 
\left(1-\sum_{\beta} (p_{i-1}^{\beta}+p_i^{\beta})+
\sum_{\gamma,\delta} \Gamma_i^{\delta\gamma}
\right)
\left(1-\sum_{\beta} (p_i^{\beta}+p_{i+1}^{\beta})+
\sum_{\gamma,\delta} \Gamma_{i+1}^{\delta\gamma}
\right)}\,.
\label{pottsstruc-eq}
\end{equation}
By solving this equation numerically we obtain the exact density
profiles. (For the example of eq. (15), results are shown in Fig.~1.)

\subsection{Mean spherical approximation (MSA)}

In the following the excess density functional will be expanded around
a homogeneous reference state, $p_i^{\alpha}=\bar p_\alpha$
\cite{refstate-comm}. For simplicity, we here consider the standard
Potts case $v_i^{\alpha\beta}=V\delta_{\alpha\beta}$, and set $\bar
p_\alpha=\bar p$ for all $\alpha$.

Defining $\Delta p_i^{\alpha}=p_i^{\alpha}-\bar p$,
$\Delta\Omega[\mb{p}]=\Omega[\mb{p}]-\Omega[\{\bar p\}]$, etc.,
we can write
\begin{equation}
  \label{d-gfe}
  \Delta \Omega[\mb{p}]=\Delta F_{\rm id}[\mb{p}]+ \Delta F_{\rm
  ex}[\mb{p}] + \sum_{i,\alpha}\tilde\epsilon_i^{\alpha}\Delta
  p_i^{\alpha}\,.
\end{equation}
When retaining only terms up to second order in the excess free energy
functional, we find
\begin{equation}
  \label{d-fex}
  \Delta F_{\rm ex}[\mb{p}]=-\sum_{i,\alpha}{c^{(1)}(\bar p) \Delta
  p_i^{\alpha}}-\frac{1}{2} \sum_{i,\alpha,j,\beta}{c^{(2)}_{\alpha
  \beta}(|i-j|,\bar p) \Delta p_i^{\alpha}\Delta p_j^{\beta}}\,,
\end{equation}
where $c^{(1)}(\bar p)$ can be subsumed into the chemical potential.
For our choice of interactions with translational symmetry,
$g_{ij}^{\alpha\beta}=g_{\alpha\beta}(|i-j|)$, and the direct
correlation function $c^{(2)}_{i\alpha,j\beta}$ can be split into two
parts,
\begin{equation}
  \label{dc}
  c_{i\alpha,j\beta}^{(2)}=
\delta_{\alpha\beta}c_1(|i-j|)+(1-\delta_{\alpha\beta})c_{2}(|i-j|)\,.
\end{equation}
where in the MSA approximation,
\begin{equation}
  \label{msa}
  c_1(|l|)=\begin{cases}
    - V,& |l|=1\\
    0,& |l|\ge 2
  \end{cases}
\hspace{1cm}\text{and}
\hspace{1cm}  c_{2}(|l|)=0 \quad \text{for} \quad |l|\ge 1
\end{equation}
and $g_{\alpha\beta}(l=0)=0$. The two unknowns $c_1(0)$ and
$c_{2}(0)$ are found using the Ornstein--Zernike
equation (\ref{eq:oz}), which yields
\begin{eqnarray}
c_1^{(2)}(0)&=& \frac{1+\bar p-q\bar p}{\bar p(1-q\bar p)}-\frac{1}{q}\left[A+(q-1)B\right]\nonumber\\
c_{2}^{(2)}(0)&=& \frac{1}{1-q\bar p}-\frac{1}{q}\left[A-B\right]\nonumber\\
 \text{with} \hspace{1cm} A&=&\frac{\sqrt{1+[2 V \bar p(1-q\bar
    p)]^2}}{\bar p(1-q\bar p)}, \; B=\frac{\sqrt{1+(2 V \bar p)^2}}{\bar p}\,.
\end{eqnarray}

As an example, we now consider the special case of zero external
energies for the Potts states $\alpha=2,\ldots,q$ and alternating
external energies for the Potts state $\alpha=1$,
\begin{equation}
\epsilon_i^{\alpha}=\begin{cases}
\epsilon,& i \; \text{even and} \quad \alpha=1\\
0,& \text{otherwise}
\end{cases}
\end{equation}
and assume $\mu_{\alpha}=\mu$ independent of $\alpha$. Due to the
symmetry of this external potential, each occupation number
$p_i^{\alpha}$ is equal to one of the four representatives
$p_0^1,p_0^{2},p_1^1,p_1^{2}$, and the mean particle density per site
is
\begin{equation}
  \label{nbar}
\bar{n} = \frac{1}{2}\left[ p_0^1+p_1^1+(q-1)(p_0^{2}+p_1^{2}) \right].  
\end{equation}

We consider the functional
\begin{equation}
\begin{split}
  \label{functional}
  \Psi [\mb{p}]  \equiv& \frac{1}{M}\left( \Delta \Omega[\mb{p}]+
  F_{\rm id}[\bar{p}] \right)\\
  =& \frac{1}{2}\bigl[p_0^1 \ln p_0^1 + p_1^1 \ln p_1^1 + (q-1)
  \left( p_0^{2} \ln p_0^{2} + p_1^{2} \ln p_1^{2}\right)\\
  &\phantom{\frac{1}{2}\bigl[}
  +\left(1-p_0^1-(q-1)p_0^{2}\right) \ln
  \left(1-p_0^1-(q-1)p_0^{2}\right)\\
  &\phantom{\frac{1}{2}\bigl[}
  +\left(1-p_1^1-(q-1)p_1^{2}\right) \ln
  \left(1-p_1^1-(q-1)p_1^{2}\right)
 \bigr]\\
&-\frac{1}{4}\biggl[ 
c_1(0) \biggl((\Delta p_0^1)^{2} + (\Delta p_1^1)^{2} +
  (q-1)\left((\Delta p_0^{2})^{2} +(\Delta p_1^{2})^{2} \right) \biggr)\\
&\phantom{-\frac{1}{4}\biggl[}
+c_{2}(0)(q-1)\left\{2\left(\Delta p_0^1\Delta p_0^{2}+\Delta p_1^1\Delta p_1^{2}\right)\right.\\
&\phantom{-\frac{1}{4}\biggl[+c_{2}(0)(q-1)}+(q-2)\left.\left((\Delta p_0^{2})^{2} + (\Delta p_1^{2})^{2}\right) \right\}\\
&\phantom{-\frac{1}{4}\biggl[}
-4 V\left( \Delta p_0^1 \Delta p_1^1 + (q-1)\Delta p_0^{2} \Delta p_1^{2}\right)
\biggr]\\
&+\frac{1}{4}\epsilon
\left[p_0^1-p_1^1-(q-1)(p_0^{2}+p_1^{2})+2(q-1)\bar{p}
  \right]\\
&-\mu_{\rm eff}
\left[\frac{1}{2}\big(p_0^1+p_1^1+(q-1)(p_0^{2}+p_1^{2}) \big)
  -\bar{n}\right]\,.
\end{split}
\end{equation}
with $\mu_{\rm eff}=\mu-\epsilon/2+ c^{(1)}(\bar{p})$. The structure
equations are obtained by setting the derivatives of $\Psi[\mb{p}]$
with respect to $p_0^1,p_0^{2},p_1^1,p_1^{2}$ equal to zero. These
equations are solved numerically subject to a fixed $\bar{n}$ in
eq.~(\ref{nbar}) (which is equivalent to $\D\Psi/\D\mu_{\rm eff}=0$).
Ordinary mean field theory can be recovered from the functional by setting
$c_1(0)=c_{2}(0)=0$.

\begin{figure}
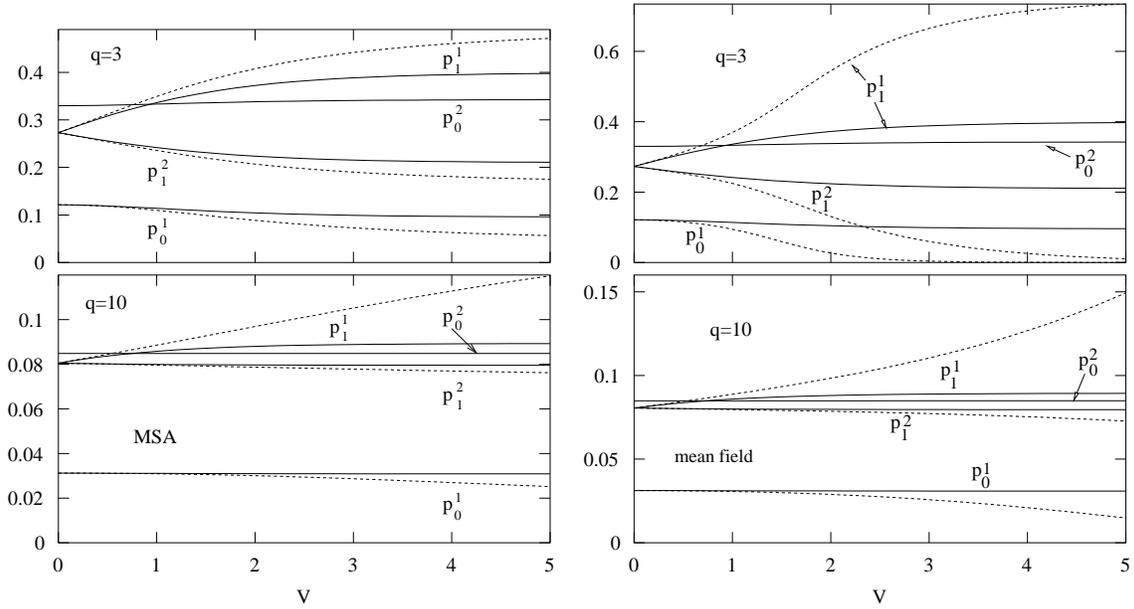

\label{msa-fig}
\begin{center}
\begin{minipage}[t]{7.2cm}
\epsfig{file=./figures/fig1a.eps,width=7.3cm}
\end{minipage}
\hspace*{0.2cm}
\begin{minipage}[t]{7.2cm}
\epsfig{file=./figures/fig1b.eps,width=7.3cm}
\end{minipage}
\end{center}
\caption{Comparison of the occupation numbers $p_0^1$, $p_1^1$ and
$p_1^2$ in dependence of interaction strength $V$ for two $q$ values
in the MSA approximation (figures on left side, dashed lines) and mean
field approximation (figures on right side, dashed lines) with the
exact solution (solid lines) for $\bar n=0.8$ and $\epsilon=1$. The
corresponding MSA and mean field results for $p_0^2$ (not shown) are
always close to the exact solution.}
\end{figure}

In Fig.~1 the occupation numbers for the MSA approximation and the
mean field theory are compared with the exact results based on eq.
(\ref{pottsstruc-eq}). Without interactions ($V=0$) the external
potential yields mean occupation numbers
$p_0^1<p_1^1=p_1^{2}<p_0^{2}$, $p_0^1=e^{-\epsilon}p_0^{2}$. When a
repulsive interaction $V>0$ is switched on, the fact that
$p_1^1>p_1^{2}$ induces an increasing occupation difference. For
larger $q$ this effect becomes less pronounced due to the increasing
contribution of the entropy to the free energy per site.

As expected, the MSA is an improvement over the simple mean field
approximation especially for higher values of the interaction
parameter and high overall densities $\bar n$. Moreover, the quality
of the MF-results can be shown to improve when $\epsilon$ is
decreased. A notable result is the improvement of the two
approximations for larger $q$, reflecting the fact that mean-field
descriptions should become exact in the limit $q \rightarrow \infty$.

\subsection{Kinetics of density profiles}

So far we have considered density profiles at equilibrium. In order to
account for the time evolution of non-equilibrium profiles we will
make use of the local equilibrium approximation for the probability
distribution $P(\mb{x},t)$ to find a configuration
$\mb{x}=\{x_i^{\alpha}\}$ at time $t$. In this time-dependent density
functional theory (TDFT)
\cite{Reinel/Dieterich:1996,Gouyet/etal:2003}, the deviations of
$P(\mb{x},t)$ from the Boltzmann equilibrium distribution are
described by a one-particle time-dependent effective potential
$h_i^{\alpha}(t)$,
\begin{equation}
P(\mb{x},t)=\frac{1}{Z(t)}\exp\left[-H(\mb{x})-\sum_{i,\alpha}
h_i^{\alpha}(t)x_i^{\alpha}\right]\,.
\label{eq:prob-non-eq}
\end{equation}
The effective potential is the unique potential, which yields the
instantaneous density profile $p_i^{\alpha}(t)$ according to the
equilibrium DFT. Accordingly, the unknown field
$\mb{h}$ in eq.~(\ref{eq:prob-non-eq}) can be determined by the
``structure equation'' $h_i^{\alpha}(t)=-\D \Omega[\mb{p}]/\D
p_i^{\alpha}$ with $\Omega[\mb{p}]$ from eq.~(\ref{eq:omega}). It is
then clear that in this approach all equilibrium relations between
occupational correlators $\langle
x_i^{\alpha}x_j^{\beta}\ldots\rangle$ and the density $\mb{p}$ also
apply at each time instant to the non-equilibrium situation [in
particular eq.~(\ref{pottscorrel-eq})].

To test the quality of the TDFT, we consider a lattice with all sites
being occupied ($\bar n=1$) and a nonconserved dynamics, where a
given Potts state $\alpha$ on lattice site $i$ can change to any other
state $\beta\ne\alpha$ with a rate $w_{i}^{\alpha\beta}(\mb{x})$ that
depends on the current state $\mb{x}$ due to interactions. From the
master-equation describing this stochastic process, we derive the
equation of motion for the occupation profile,
\begin{equation}
\frac{d p_i^{\alpha}}{dt}=\sum_\beta \langle
(x_i^{\beta}-x_i^{\alpha})w_{i}^{\alpha\beta}\rangle_t
\label{eq:rate-me}
\end{equation}
where $\langle\ldots\rangle_t$ denotes an average with respect to the
non-equilibrium distribution $P(\mb{x},t)$. For the rates we choose a
generalized Glauber form,
\begin{eqnarray}
w_{i}^{\alpha\beta}(\mb{x})&=&
\left(1-\tanh\Bigl[(x_{i-1}^{\beta}x_i^{\beta}-
                    x_{i-1}^{\alpha}x_i^{\alpha})V/2\Bigr]\right)\nonumber\\
&&{}\times\left(1-\tanh\Bigl[(x_{i+1}^{\beta}x_i^{\beta}-
                              x_{i+1}^{\alpha}x_i^{\alpha})V/2\Bigr]\right)\,,
\label{eq:rate-tanh}
\end{eqnarray}
which satisfies the condition of detailed balance \cite{rate-comm}.

When substituting eq.~(\ref{eq:rate-tanh}) in (\ref{eq:rate-me})
further evaluation is made possible by noting that the factor in front
of $V/2$ in the argument of $\tanh$ is always $-1,0$ or $1$, so that
it can be taken out from the $\tanh$ function. With the help of the
Markov property, three-point correlators can exactly be reduced to
two-point correlators, e.g.\ $\langle x_{i-1}^{\alpha} x_i^{\beta}
x_{i+1}^{\gamma} \rangle= \Gamma_i^{\alpha\beta}
\Gamma_{i+1}^{\beta\gamma} / p_i^{\beta} $. After some algebra we then
find
\begin{eqnarray}
\frac{d p_i^{\alpha}}{dt}&=&
1-q p_i^{\alpha} + \tanh(V/2) \Bigl[2p_{i}^{\alpha}-p_{i-1}^{\alpha}-p_{i+1}^{\alpha}\nonumber\\
&&\hspace*{5em}{}+\sum_\beta\Bigl(
  \Gamma_{i}^{\beta\beta}-\Gamma_{i}^{\alpha\alpha}
 +\Gamma_{i+1}^{\beta\beta}-\Gamma_{i+1}^{\alpha\alpha}
\Bigr) \Bigr]\nonumber\\
&&{}+\tanh^2(V/2)\sum_\beta\Bigl[\frac{1}{p_i^{\beta}} 
    (\Gamma_{i}^{\alpha\beta}-\Gamma_{i}^{\beta\beta})
    (\Gamma_{i+1}^{\beta\alpha}-\Gamma_{i+1}^{\beta\beta})\nonumber\\
&&\phantom{\tanh(V/2)^2\sum_\beta\Bigl[}
{}-\frac{1}{p_i^{\alpha}}
    (\Gamma_{i}^{\beta\alpha}-\Gamma_{i}^{\alpha\alpha})
    (\Gamma_{i+1}^{\alpha\beta}-\Gamma_{i+1}^{\alpha\alpha})
\Bigr].
\label{eq:tdft-time-evol}
\end{eqnarray}

Equations (\ref{eq:tdft-time-evol}) together with the correlator equations
(\ref{pottscorrel-eq}) form a complete set of equations for the time
evolution of density profiles and can be solved numerically for a
given initial condition.

\begin{figure}
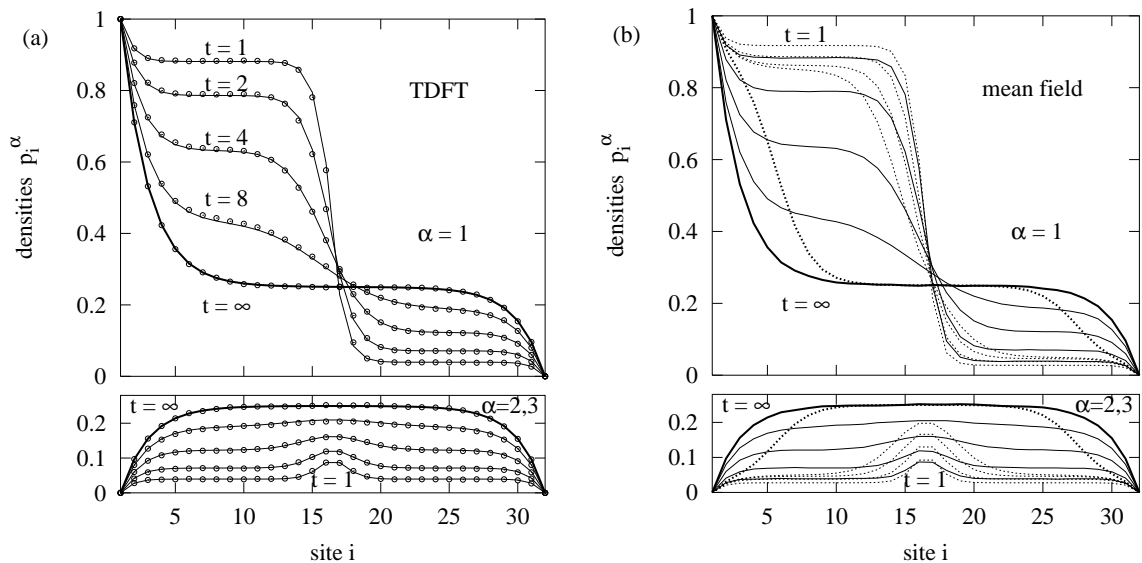

\begin{center}
\begin{minipage}[t]{7.2cm}
\epsfig{file=./figures/fig2a.eps,width=7.3cm}
\end{minipage}
\hspace{0.4cm}
\begin{minipage}[t]{7.2cm}
\epsfig{file=./figures/fig2b.eps,width=7.3cm}
\end{minipage}
\end{center}
\caption{(a) Comparison of the evolution of an initial kink profile
calculated using TDFT (lines) with Monte Carlo simulations (points)
for a system with 32 particles, $q=4$ and $V=-2$. Boundary conditions
are $p_1^\alpha = \delta_{\alpha,1}$ and $p_M^\alpha =
\delta_{\alpha,4}$. Results are shown for $t=1,2,4,8$ and the
equilibrium density (thick lines). Results for $\alpha=4$ (not shown)
are mirror images of those for $\alpha=1$ with respect to the center
site $i=16$. (b) Comparison of mean field theory (dotted lines) with
the simulations (solid lines) for the same system as shown in (a).
Equilibrium solutions are indicated with thick lines.}
\label{fig:tdft-mc-test}
\end{figure}

As an example, we consider an initially sharp kink profile at time
$t=0$, where the left part of the system is in Potts state $\alpha=1$
and the right part in Potts state $\alpha=q$ with fixed boundary sites
$i=1$ and $M$ in Potts states $\alpha=1$ and $q$, i.e.\ 
$x_1^{\alpha}=\delta_{\alpha,1}$ and
$x_{M}^{\alpha}=\delta_{\alpha,q}$. In Figure~\ref{fig:tdft-mc-test}a
we compare the time evolution of the profile with the results from
continuous-time Monte-Carlo simulations. The agreement is excellent
for all times until for large times the TDFT solution and the
simulations both yield the correct equilibrium profile. In order to
obtain this agreement, it was necessary to adjust the time scale
according to $t_{\rm TDFT}=0.85\; t_{\rm MC}$. The significant
improvement over a simple mean-field treatment corresponding to
factorization of all correlators in eq.(\ref{eq:tdft-time-evol}),
i.e.\ $\Gamma_i^{\alpha\beta}=p_{i-1}^{\alpha}p_i^{\beta}$, etc., can
be seen by comparing Fig.~\ref{fig:tdft-mc-test}a with
Fig.~\ref{fig:tdft-mc-test}b. The mean-field approximation is
insufficient from the beginning and does not provide the correct
equilibrium profile for long times.

\subsection{Basic Limitations of TDFT}
Since in the kinetic equation of the TDFT only the densities (mean
occupation numbers) enter, the TDFT can not distinguish between states
with the same density profile but different correlations. As a
consequence, the time evolution of a non-equilibrium state with the
equilibrium density profile cannot be captured. This issue may become
important when considering memory effects in systems with slow
relaxations, as for example glassy systems. The observables used to
characterize the thermodynamic states of such systems (as e.g.\
density) may have equal values initially, but very different time
evolutions depending on the systems' history.

To illustrate this principal failure of the TDFT, we generate states
of the Potts model with the same mean occupation numbers but different
correlations for the same system as in the previous chapter. To
achieve this, we first determine the equilibrium profile
$p_i^{\alpha}$ for a system with interactions from the structure
equations (\ref{pottsstruc-eq}). This profile is also the equilibrium
profile for a system without interactions ($V=0$) and external
potential $\epsilon_i^{\alpha}=-\log p_i^{\alpha}$. Hence by taking
the equilibrium state of the non-interacting system as initial
non-equilibrium state for the interacting system, we can study by
Monte-Carlo simulations the time development of a density profile that
according to TDFT cannot change.

\begin{figure}
\begin{center}
\epsfig{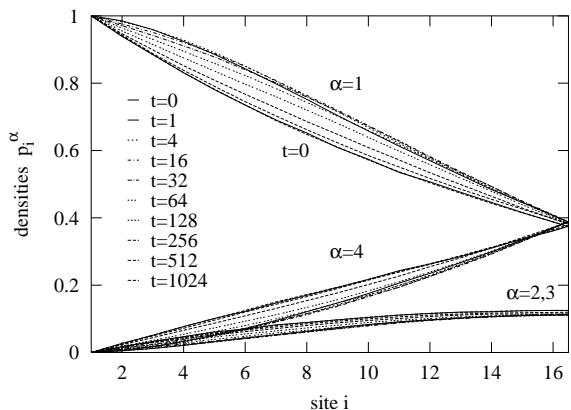}
\end{center}
\caption{Evolution of a density profile starting from the equilibrium 
  density, but with non-equilibrium correlators (see text) for $V=-4$
  and $q=4$.  The system quickly reaches the state with maximum
  deviation from the equilibrium density at $t\approx 1$ and then
  relaxes to the original equilibrium density profile. Only the left
  half of the system is shown.}
\label{fig:tdft-failure}
\end{figure}

Figure~\ref{fig:tdft-failure} shows the effect for $V=-4$ and $q=4$.
Although the mean occupation numbers at time $t=0$ are the same as in
the equilibrium state reached for $t\simeq1024$, there is a pronounced
change in the occupation profile at intermediate times. These changes
are due to the fact that in the initial state the correlations between
occupation numbers are not the equilibrium ones. The configurations
with large statistical weights in the initial state exhibit stronger
short-wavelength fluctuations and in order for the correlations to
build up, the system has to pass through intermediate states with a
non-equilibrium profile.

\subsection{Acknowledgments}
We gratefully acknowledge financial support by the
Sonderforschungsbereich 513 of the Deutsche Forschungsgemeinschaft.

%---------------------------------------------------------------------------


\begin{thebibliography}{9}
\bibitem{refstate-comm} An alternative possibility would be to choose
  a reference state being homogeneous in space but inhomogeneous in
  the space of Potts states, i.e.\ $p_i^{\alpha}=\bar p_\alpha$.  In
  certain applications this might be advantageous. For simplicity,
  however, we will not consider this case here.
\bibitem{rate-comm} For a Hamiltonian $H$ that is decomposable into
  independent parts, $H=\sum_i H_i$, the generalized Glauber rate
  $w_{ab}$ for a transition from an initial state $a$ to a final state
  $b$ in the form $w_{ab}=\prod_i
  \{1-\tanh[(H_i^{(b)}\!-\!H_i^{(a)})/2]\}$ always satisfies the
  condition of detailed balance.
\end{thebibliography}
\end{document}